\documentclass[journal, 10pt,twocolumn]{IEEEtran}

\usepackage{caption}

\usepackage{ifdraft}
\usepackage{lipsum} 
\usepackage{bbm}
\usepackage{booktabs}
\usepackage{pdfcomment}
\usepackage{subcaption}
\usepackage{graphicx}
\usepackage{color,soul}
\usepackage{amsthm} 
\theoremstyle{plain}

\theoremstyle{definition}

\usepackage{cite}
\usepackage{amsmath,amssymb,amsfonts}
\usepackage{algorithm}
\usepackage{algorithmic}
\usepackage{graphicx}
\usepackage{textcomp}
\usepackage{xcolor} 
\usepackage{xpatch}
\makeatletter
\ExplSyntaxOn
\cs_new:Npn \bibColoredItems #1#2
{
\clist_map_inline:nn {#2} { \cs_new:cpn {bib@colored@##1} {#1} } 
}
\ExplSyntaxOff

\newcommand\bib@setcolor[1]{%
\ifcsname bib@colored@#1\endcsname
\expanded{\noexpand\color{\csname bib@colored@#1\endcsname}}%
\else
\normalcolor
\fi
}

\IfPackageLoadedTF{hyperref}{\@tempswatrue}{\@tempswafalse}
\if@tempswa
\xpatchcmd\@bibitem {\H@item}{\bib@setcolor{#1}\H@item}{}{\PatchFailed}
\xpatchcmd\@lbibitem{\H@item}{\bib@setcolor{#2}\H@item}{}{\PatchFailed}
\else
\xpatchcmd\@bibitem {\item} {\bib@setcolor{#1}\item} {}{\PatchFailed}
\xpatchcmd\@lbibitem{\item} {\bib@setcolor{#2}\item} {}{\PatchFailed}
\fi
\makeatother

\usepackage{color}
\usepackage{bm}
\usepackage{booktabs}
\usepackage{cleveref}

\usepackage{geometry}
\ifCLASSINFOpdf

\else

\fi

\usepackage{changes} 
\begin{document}
\title{\color{black}Sparse Recovery for Holographic MIMO Channels: Leveraging the Clustered Sparsity}

\author{
Yuqing Guo, 
Xufeng Guo,
Yuanbin Chen,~and~Ying~Wang,~\IEEEmembership{Member,~IEEE}
\vspace{-0.5cm}
}

\maketitle
\begin{abstract}
Envisioned as the next-generation transceiver technology, the holographic multiple-input-multiple-output (HMIMO) garners attention for its superior capabilities of fabricating electromagnetic (EM) waves.
However, the densely packed antenna elements significantly increase the dimension of the HMIMO channel matrix, rendering traditional channel estimation methods inefficient. 
While the dimension curse can be relieved to avoid the proportional increase with the antenna density using the state-of-the-art wavenumber-domain sparse representation, the sparse recovery complexity remains tied to the order of non-zero elements in the sparse channel, which still considerably exceeds the number of scatterers.
By modeling the inherent clustered sparsity using a Gaussian mixed model (GMM)-based von Mises–Fisher (vMF) distribution, the to-be-estimated channel characteristics can be compressed to the scatterer level.
Upon the sparsity extraction, a novel wavenumber-domain expectation-maximization (WD-EM) algorithm is proposed to implement the cluster-by-cluster variational inference, thus significantly reducing the computational complexity.
Simulation results verify the robustness of the proposed scheme across overheads and signal-to-noise ratio (SNR).



\end{abstract}
\begin{IEEEkeywords}
Holographic MIMO, channel estimation, sparse recovery, wavenumber domain, clustered sparsity
\end{IEEEkeywords}

\section{Introduction}\label{sec4}
Holographic multiple-input-multiple-output (HMIMO) has been envisioned as the next-generation transceiver technology in future 6G communications~\cite{2024TieruiGongJSAC}. Specifically, HMIMO is capable of achieving super-directivity~\cite{2019MarzettaACSSC} and theoretically unlimited spectral efficiency~\cite{PartIII}, facilitated by its spatially continuous aperture, which incorporates a large number of radiation elements densely arranged on a meta-material-based surface \cite{RHS}. 
As a prerequisite for realizing these potential applications, acquiring accurate channel state information (CSI) in advance is essential.

However, unlike traditional MIMO systems, the spacing between antenna elements in HMIMO is significantly less than half the wavelength. This results in the dimensions of the channel matrix growing quadratically for the same array aperture, which causes the traditional estimation of the entire channel matrix to fail due to the curse of dimensionality.
To address the significantly increased dimensionality in channel estimation, an effective strategy is to exploit the inherent sparsity of channels within the wireless propagation environment. Specifically, estimating only the non-zero elements of the sparse channel representation substantially reduces the dimensionality of the parameters required for estimation. A classical method for achieving this is by transforming channels into their sparse angular-domain counterparts using discrete Fourier transformation (DFT)~\cite{AD}. In high-frequency channels, electromagnetic (EM) waves typically travel only through a limited number of paths corresponding to scatterers in the EM field. Therefore, robust channel recovery can be efficiently achieved by focusing the estimation on these non-zero elements that are associated with the scatterers in the angular-domain sparse representation.

Nevertheless, the sparsity represented in the angular-domain sparse representation achieves optimal performance only under the ideal assumptions of half-wavelength spaced antennas and far-field plane wave propagation~\cite{Polar1}. In HMIMO systems, where the antenna spacing theoretically approaches infinitesimally small, continued use of traditional angular-domain methods can lead to issues of power leakage and probe redundancy~\cite{gxfICC}. To address these challenges, the state-of-the-art wavenumber-domain sparse channel representation can be adopted. This method effectively represents continuous electromagnetic propagation waves using a finite set of orthogonal and linear Fourier harmonics with dimensions only related with the aperture size~\cite{fourier}.
By representing the HMIMO channel in the wavenumber domain, we can exploit the benefits of structured sparsity to refine the robustness of sparse channel recovery. Specifically, the locations of non-zero elements in the sparse representation typically exhibit a particular structure, often clustering together, referred to as the clustered sparsity. In~\cite{RobustRecovery}, a Markov chain-based model is proposed to capture the clustered sparsity, demonstrating outstanding performance in uniform linear array (ULA) systems. An extended exploitation of clustered sparsity in the uniform planar array (UPA) system using a Markov field-based model is posed in~\cite{gxfICC}. Furthermore, the clustered sparsity can be leveraged in the three-dimensional free space through three joint Markov chain modeling in~\cite{vtguo}.

However, previous work on clustered sparsity has exclusively focused on the behaviors of the locations of non-zero elements while neglecting the behaviors of the values of non-zero elements within clustered sparsity. According to the Gaussian mixed model (GMM)-based von Mises–Fisher (vMF) distribution~\cite{spatialcharacter}, the values corresponding to elements closer to the cluster center are more significant than those farther from the center. On the other hand, these approaches all rely on Markov process modeling, whose complexity is still proportional to the number of non-zero elements in the wavenumber domain. Theoretically, according to the vMF distribution, the channel characteristics are determined by a finite number of propagation paths corresponding to scatterers, the number of which is significantly less than that of the non-zero elements.

\begin{figure*}
    \centering
    \begin{minipage}{.32\textwidth}
    \centering
        \includegraphics[width=0.9\linewidth]{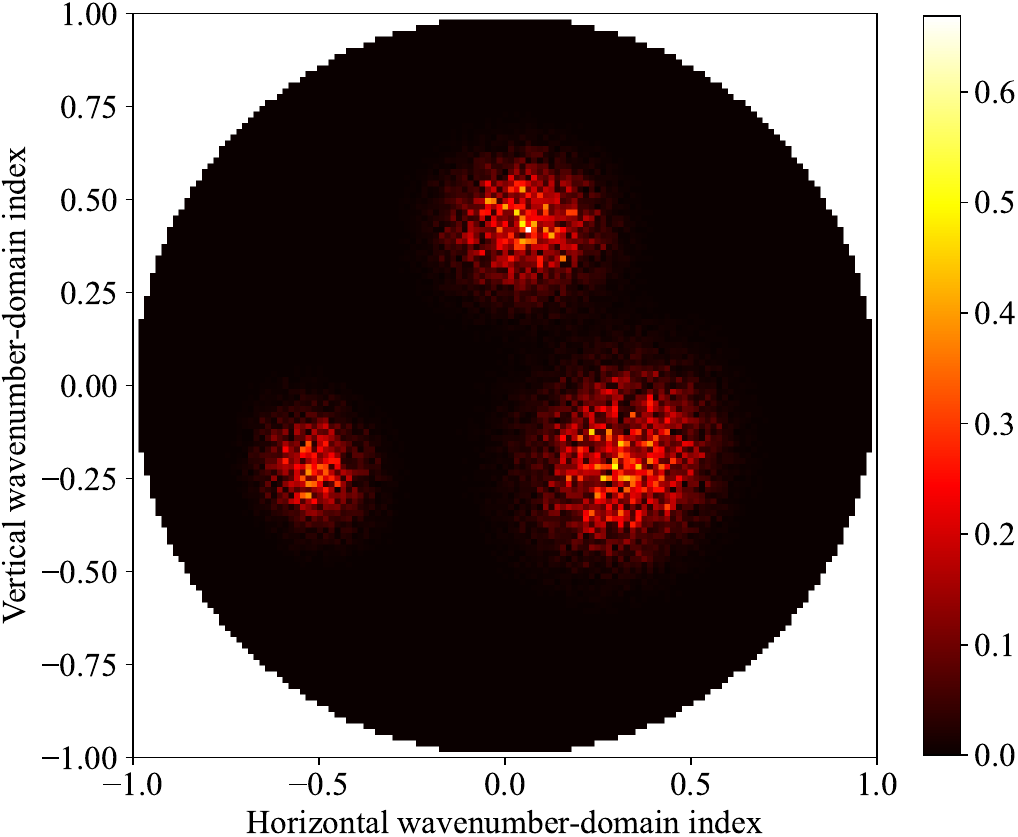}
        \subcaption{Observed snapshot of the HMIMO channel on the wavenumber domain.}
    \end{minipage}
    \hfill
    \begin{minipage}{.32\textwidth}
    \centering
        \includegraphics[width=0.78\linewidth]{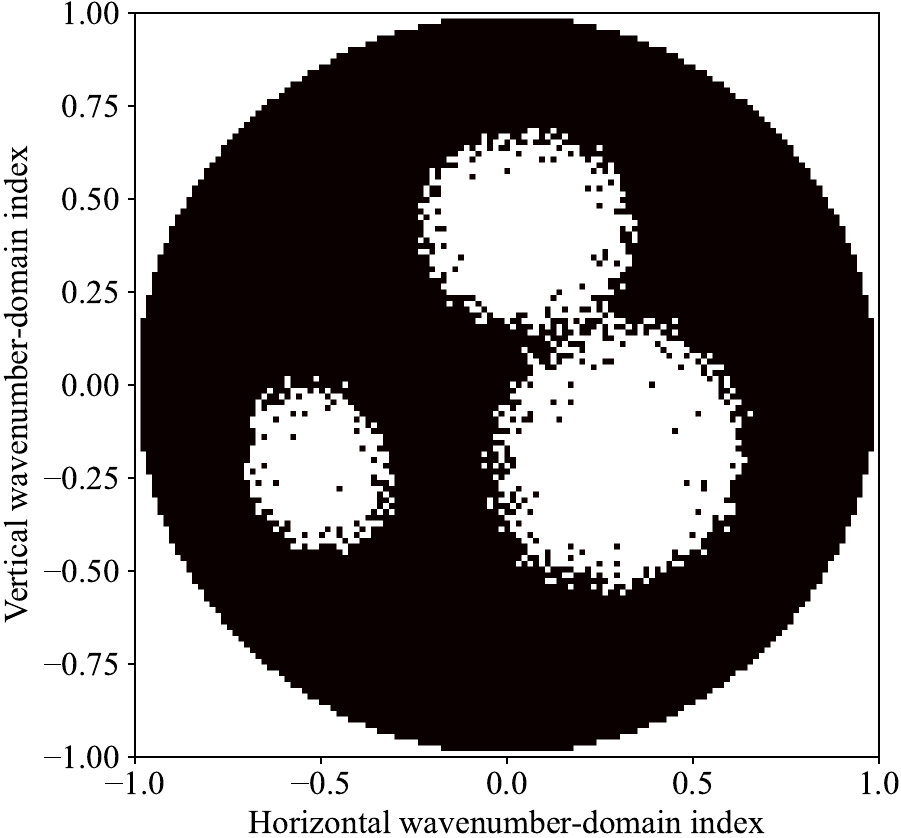}
        \subcaption{Traditional sparse recovery methods estimate all the non-zero entries.}
    \end{minipage}
    \hfill
    \begin{minipage}{.32\textwidth}
    \centering
        \includegraphics[width=0.78\linewidth]{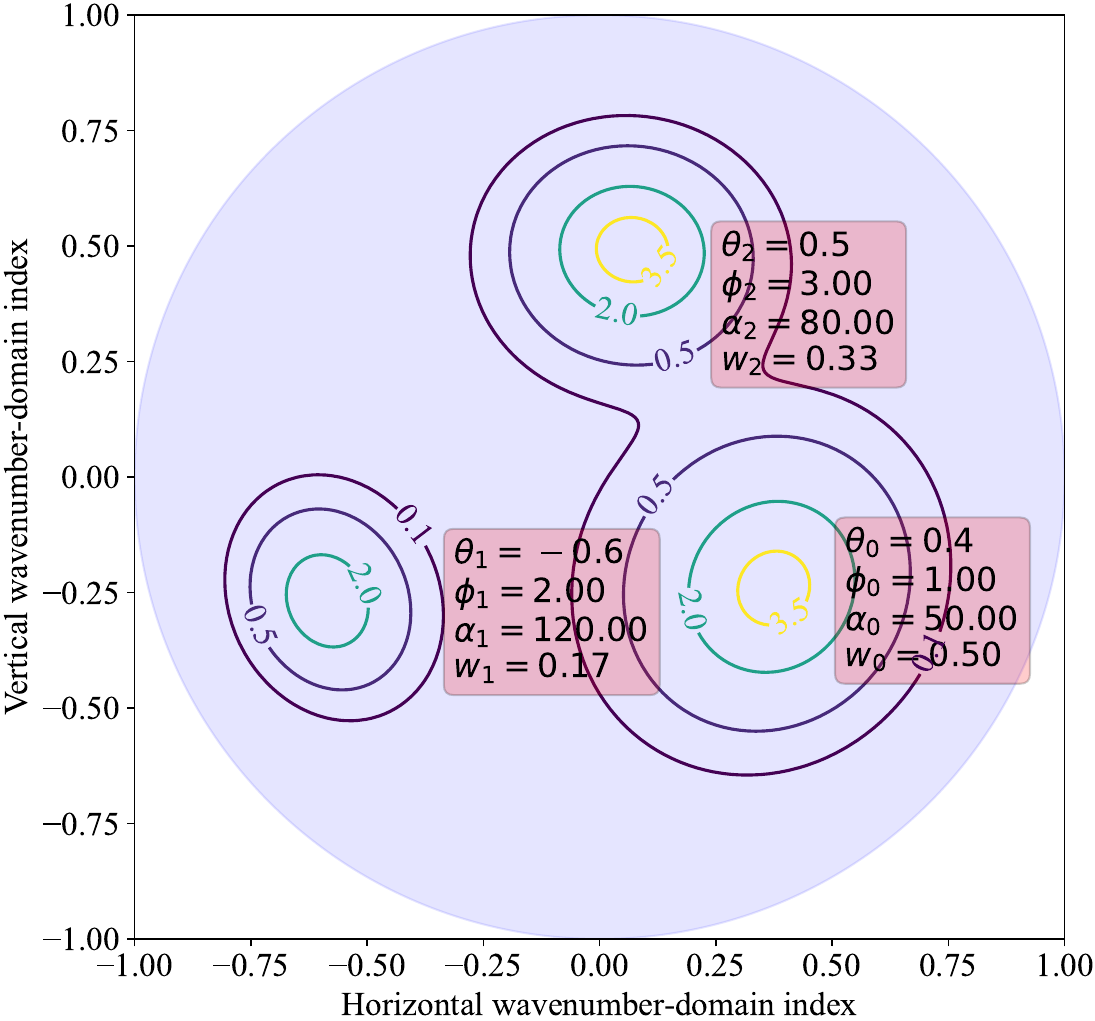}
        \subcaption{Our proposed scheme aims to estimate only the GMM-based parameters.}
    \end{minipage}
    \caption{Illustration of our proposed sparse recovery facilitated by the clustered sparsity.}
    \label{fig:illustration}
    \vspace{-0.5cm}
\end{figure*}

This paper explores wavenumber-domain sparse channel recovery, facilitated by clustered sparsity, for HMIMO systems. Our contributions are summarized as follows:
\begin{itemize}
    \item Firstly, we transform the high-dimensional HMIMO channel estimation problem into a variational inference problem in the wavenumber domain, characterized by a GMM-based von Mises–Fisher (vMF) distribution, with the number of parameters to be estimated being on the order of the number of scatterers.
    \item Secondly, a wavenumber-domain expectation-maximization (WD-EM) algorithm is constructed to find the most likely estimates of the GMM parameters with extremely low computational complexity.
    \item Finally, numerical simulations confirm the robustness of the proposed WD-EM algorithm against both the number of observations and the signal-to-noise ratio (SNR).
\end{itemize}

\section{System Model}\label{sec2}
\subsection{Signal Model}
We consider the uplink multiple-input-multiple-output (MIMO) communication system where the base station (BS) is equipped with Holographic MIMO (HMIMO), serving several single-antenna users.
The signals used in communications are narrow-band with central frequency $f_c$. The HMIMO is parallel with $xOy$-plane and equipped with $N=N_x\times N_y$ antenna elements with antenna spacing $\delta=\lambda/2$, spanning the rectangular region $\mathcal{R}\triangleq[0,N_x\delta]\times[0,N_y\delta]$. For simplicity, we assume that each user sends a mutually orthogonal pilot sequence to the BS for channel estimation so that the channel estimation for each user is independent. At any specific symbol time, the received signal at the BS sent by an arbitrary user can be expressed as:
\begin{equation}\label{yk}
\mathbf{y}=\mathbf{C}\mathbf{H}x+\mathbf{n},
\end{equation}
where $\mathbf{H}\in\mathbb{C}^{N\times1}$ denotes the uplink channel between the user and the HMIMO; $\mathbf{C}\in\mathbb{C}^{N_{\rm RF}\times N}$ denotes the combining matrix which lowers the dimension of the signal from $N$ to $N_{\rm RF}$; $\mathbf{y}\in\mathbb{C}^{N_{\rm RF}\times1}$ is the received signal at the RF chain; $x\in\mathbb{C}$ denotes the symbol sent by the user; $\mathbf{n}\in\mathbb{C}^{N_{\rm RF}\times1}$ denotes the additive white Gaussian noise (AWGN) with covariance matrix $\mathbf{n}\sim\mathcal{CN}(0,\sigma_0^2\mathbf{I}_{N_{\rm RF}})$.
\subsection{Sparse Channel Model}
We suppose the BS lies on the plane $z=z_0$ and the user lies on the plane $z=0$. 
According to~\cite{fourier}, the channel response from the user to the plane $z=z_0$ can be modeled as the superposition of a finite set of planar waves termed the Fourier harmonics (FHs).
Thus, the channel response $h(r_x,r_y)$ on the arbitrary point $(r_x, r_y)\in \mathcal{R}$ on the HMIMO array is given by
\begin{equation}\small
\begin{aligned}
\label{hxy}
&h(r_x,r_y)=
\sum_{(m_x,m_y)\in\xi}H(m_x,m_y)\exp \Big\{-j\Big[\frac{2\pi m_x}{L_x}r_x
\\
&+\frac{2\pi m_y}{L_y}r_y+\sqrt{(\frac{2\pi f_c}{c})^2-(\frac{2\pi m_x}{L_x})^2-(\frac{2\pi m_y}{L_y})^2}z_0\Big]\Big\},
\end{aligned}
\end{equation}
where $L_x=N_x\delta$ and $L_y=N_y\delta$ denote the HMIMO aperture along the $x$- and $y$-axis, respectively. 
$\xi$ is the index set of the FHs that satisfies the following restriction to exclude the evanescent waves that can be neglected since they decay exponentially along the $z$-axis:
\begin{equation}\label{xi}\small
\xi\triangleq\left\{(m_x,m_y)\in {\mathbb{Z}}^2:(\frac{2\pi m_x}{L_{x}})^2+(\frac{2\pi m_y}{L_{y}})^2\leq (\frac{2\pi f_c}{c})^2\right\}.
\end{equation}
$H(m_x,m_y)$ in (\ref{hxy}) denotes the weight factor of the $(m_x, m_y)$-th FH, referred to as the random Fourier coefficient, which is modeled as the complex Gaussian distribution, i.e., 
    $H(m_x, m_y)\sim \mathcal{CN}(0,\sigma^2(m_x,m_y)$,
in which the covariance factor $\sigma^2(m_x,m_y)$ is determined by the integration in the corresponding wavenumber domain:
\begin{equation}\label{eq:sigma}\small
\sigma^2\left(m_x, m_y\right)=
\int_{\frac{2 \pi m_y}{L_{ y}}}^{\frac{2 \pi\left(m_y+1\right)}{L_{ y}}} \int_{\frac{2 \pi m_x}{L_{ x}}}^{\frac{2 \pi\left(m_x+1\right)}{L_{ x}}}
\frac{A^2\left(k_x, k_y\right)}{\sqrt{k^2-k_x^2-k_y^2}} {\rm d} k_x {\rm d} k_y .
\end{equation}
By taking $k_{x}=k\sin\ \theta\cos\ \phi$, $k_{y}=k\sin\ \theta\sin\ \phi$ and $k_{z}=k\cos\ \theta$, (\ref{eq:sigma}) can be recast into:
\begin{equation}\label{sigma2}
\begin{aligned}
\sigma^2(m_x,m_y)=\iint_{\Omega(m_x,m_y)}A^2(\theta,\phi)d\theta d\phi,
\end{aligned}
\end{equation}
where $A^2(\theta,\phi)$ can be modeled as the mixture of von Mises–Fisher (vMF) distribution~\cite{vmfdistribution} under non-isotropic propagation environment:
\begin{subequations}\label{vmf}
\begin{align}
&A^2(\theta,\phi)=\sum^{N_c}_{j=1}w_jp_j(\theta,\phi), \\
&\begin{aligned}
    &p_j(\theta,\phi)=\frac{\alpha_j\sin\theta}{4\pi\sinh\alpha_j}\cdot\\ 
&\quad\exp\{\alpha_j[
       \cos\theta\cos\theta_j
      + \sin\theta \sin\theta_j\cos(\phi-\phi_j)
]\},
\end{aligned}
\end{align}
\end{subequations}
in which $N_c$ denotes the number of scatterers; $\sum_{j=1}^{N_c}w_j=1$ denote the weight factors of the scatterers; $\alpha_j$ denotes the concentration parameter of the $j$-th scatterer; $\phi^{(j)}_R$ and $\theta^{(j)}_R$ denote the azimuth angle and the zenith angle of the $j$-th scatterer, respectively. 
Recall the format in~\eqref{hxy}, the original channel matrix ${\bf H}$ can be expressed as its wavenumber-domain sparse counterpart, i.e., ${\bf G}\in \mathbb{C}^{|\xi|\times 1}$ as follows:
\begin{equation}\label{SR}
\mathbf{H}=\mathbf{\Psi}\mathbf{G},
\end{equation}
where ${\bf \Psi}\in \mathbb{C}^{N\times |\xi|}$ is interpreted as the wavenumber-domain \textit{dictionary matrix}, the $m$-th column of which denotes the corresponding channel response vector of the $m$-th FH, that is,
\begin{equation}
    [\mathbf{\Psi}]_{:,m}=\left[\frac{1}{\sqrt{N}}\exp\Big\{j\left(\frac{2\pi m_xn_{x}\delta}{L_{x}}+\frac{2\pi m_yn_{y}\delta}{L_{y}}\right)\Big\}\right]_{n\in \mathcal{N}},
\end{equation}
where $m\triangleq (m_x,m_y) \in \xi $ is defined as the two-dimensional index of the FHs.
The $m$-th entry of ${\bf G}$, i.e., $G_m \triangleq [{\bf G}]_{m}$ captures the $m$-th Fourier coefficient, i.e., $G_m = H(m_x, m_y)$.


\section{ Estimation of Statistical Channel State Information}\label{sec3}

Instead of the direct estimation of the actual channel matrix ${\bf H}$, which has the dimensionality proportional to the antenna numbers, many compressive sensing (CS)-based schemes turn to the estimates of various sparse representations of the MIMO channel.
For instance, this paper adopts the state-of-the-art wavenumber-domain sparse representation as illustrated in~\eqref{SR}, an example of the wavenumber-domain channel snapshot is given in~Fig.~\ref{fig:illustration}(a), where the traditional CS methods aim to estimate all the non-zero entries, as illustrated in~Fig.~\ref{fig:illustration}(b).
However, the dimensions of such sparse representation can still be challenging since the number of FHs, i.e., $|\xi|$ can still be extensively increased with the antenna aperture and the carrier frequency, i.e.,
$
    |\xi| \approx \left\lfloor\pi \frac{L_x L_y}{\lambda^2}\right\rfloor \propto L_x L_y \cdot f_c^2
$.
To further reduce the dimensionality, we turn to leverage the statistical characteristics of the HMIMO channel, that is, the probability density function (PDF) of the vMF distribution described in~\eqref{vmf}.
The parameter set contained in the vMF PDF is denoted as
$
    \Theta \triangleq \left\{ \omega_j, \alpha_j, \theta_j, \phi_j \right\}_{j\in \{1, \dots, N_c\}}
$.
Therefore, the number of the to-be-estimated parameters can be significantly reduced from $\mathcal{O} \left(L_xL_yf_c^2\right)$ to $\mathcal{O} \left(N_c\right)$, if we direct estimate the statistical parameters $\Theta$, as illustrated in Fig.~\ref{fig:illustration}(c).
\begin{algorithm}[t]
{\small
\caption{The proposed WD-EM algorithm}\label{algorithmchart}
{\bf Input:} Received pilot signal $\mathbf{y}$ under the combining matrix (\ref{C}); 
maximum number of scatterers $\tilde{N}_c$. 

{\bf Output:} $\{\hat{w}_j, \hat{\alpha}_j, \hat{\theta}, \hat{\phi}_j\}_{j\in \{1,\dots, \tilde{N}_c\}}$.
\begin{algorithmic}[1]\label{alg}
\STATE Initialize the parameters as $\hat{w}_j=1/\tilde{N}_c$, $\hat{\alpha}_j=1$, $\hat{\theta}_j\sim\mathcal{U}(0,\pi/2)$ and $\hat{\phi}_j\sim\mathcal{U}(0,2\pi)$.
\FOR{$i\in\{1,2,3,...,N_{\rm RF}\}$}
    \STATE Map the one-dimensional index $\lfloor\frac{\vert\xi\vert} {N_{\rm RF}}\rfloor(i-1)+1$ into its two-dimensional counterpart $(m_{x,i},m_{y,i})$ in the lattice ellipse (\ref{xi}).
    \STATE Calculate the corresponding sampling angle $(\theta_i,\phi_i)$ of the $(m_{x,i},m_{y,i})$-th wavenumber according to (\ref{thetai}) and (\ref{phii}).
    \STATE Derive the wavenumber-domain sampling values $s_i = |[{\bf y}]_i|^2$.
\ENDFOR
\WHILE{Not Converge}
\STATE \textbf{E-step:} Categorize the sampling points according to~\eqref{sj}, getting $\hat{s}_{i,j}$.
\STATE \textbf{M-step:} Given $\hat{s}_{i,j}$ from E-step to update $\hat{\phi}_j$, $\hat{\theta}_j$, $\hat{\alpha}_j$, and $\hat{w}_j$ according to \eqref{phi}, \eqref{theta}, \eqref{alpha}, and \eqref{wj}, respectively.
\ENDWHILE
\end{algorithmic}}
\end{algorithm}

\subsection{Maximum-Likelihhod Estimation of vMF Distribution}\label{MLofVMF}
Recall each scatterer's corresponding PDF in the vMF distribution shown in~\eqref{vmf}:
\begin{equation}\label{dingyi}
\begin{aligned}
    &p(\theta,\phi;\alpha_j, \theta_j, \phi_j)=
    \\
    &\frac{\alpha_j\sin\theta}{4\pi \sinh(\alpha_j)}
    \exp\{
        \alpha_j(\sin\theta\sin\theta_j\cos(\phi-\phi_j)
        +\cos\theta\cos\theta_j)
    \},
\end{aligned}
\end{equation}
in which the random variables $\theta\in(0,\pi)$ and $\phi\in(0,2\pi)$. 
Since the observed signal at the BS is $N_{\rm RF}$-dimensional, we assume that there are $N_{\rm RF}$ sampling values, each of which observes a selected point on the wavenumber domain, representing the angular direction of $(\theta_i, \phi_i), \forall i\in \{1\dots, N_{\rm RF}\}$ with the {\color{black}sampling value $s_{i,j}$.}
The sampling process is achieved by properly designing the combining matrix ${\bf C}$, as illustrated in~Sec.~\ref{CMdesign}.
Therefore, the likelihood function can be given by:
\begin{equation}\label{likelihood}
\mathcal{L}(\alpha_j,\theta_j,\phi_j)=\prod_{i=1}^{N_{\rm RF}}\{\ [p(\theta_i,\phi_i; \alpha_j,\theta_j,\phi_j))]^{s_{i,j}}\ \}.
\end{equation}
When $\mathcal{L}$ reaches its maximum value, we have:
$
\frac{\partial \log \mathcal{L}}{\partial\alpha_j}=0
$, $ 
\frac{\partial \log \mathcal{L}}{\partial\theta_j}=0
$, $
\frac{\partial \log \mathcal{L}}{\partial\phi_j}=0
$,
which are respectively equivalent to:
\begin{subequations}
    \begin{align}
        & \begin{aligned}\label{l-alpha}
            &\sum_{i=1}^{N_{\rm RF}}s_{i,j}\left\{\sin\theta_i\sin\theta_j\cos(\phi_i-\phi_j)+\cos\theta_i\cos\theta_j\right\}
            \\
            &\qquad\qquad\qquad\qquad= \left(\coth\alpha_j - \frac{1}{\alpha_j}\right)\sum_{i=1}^{N_{\rm RF}}s_{i,j},
        \end{aligned}
        \\
        & \begin{aligned}\label{l-theta}
            &\cos\theta_j\sum_{i=1}^{N_{\rm RF}}s_{i,j}\sin\theta_i\cos(\phi_i-\phi_j)
            \\
            &\qquad\qquad\qquad\qquad=\sin\theta_j\sum_{i=1}^{N_{\rm RF}}s_{i,j}\cos\theta_i,
        \end{aligned}
        \\
        & \begin{aligned}\label{l-phi}
            &\cos\phi_j\sum_{i=1}^{N_{\rm RF}}s_{i,j}\sin\theta_i\sin\phi_i
            \\
            &\qquad\qquad\qquad\qquad=\sin\phi_j\sum_{i=1}^{N_{\rm RF}}s_{i,j}\sin\theta_i\cos\ \phi_i.
        \end{aligned}
    \end{align}
\end{subequations}
By solving \eqref{l-phi} and \eqref{l-theta}, the estimates of $\{\theta_j, \phi_j\}, \forall j\in \{1,\dots, N_c\}$ can be respectively obtained by
\begin{subequations}
    \begin{align}
        \label{phi}
    \hat{\phi}_j&= \begin{cases}\arctan ({\mu_1}/{\eta_1}), & \text { if } \mu_1>0, \eta_1>0, \\ \arctan ({\mu_1}/{\eta_1})+\pi, & \text { if } \eta_1<0, \\ \arctan ({\mu_1}/{\eta_1})+2 \pi, & \text { if } \mu_1<0, \eta_1>0,\end{cases}
    \\
    \label{theta}
    \hat{\theta}_j&= \begin{cases}\arctan ({\mu_2}/{\eta_2}), & \text { if } \mu_2 \eta_2>0, \\ \arctan ({\mu_2}/{\eta_2})+\pi, & \text { if } \mu_2 \eta_2<0,\end{cases}
    \end{align}
\end{subequations}
where $\{\mu_1, \eta_1\}$ and $\{\mu_2, \eta_2\}$ are the substitution variables given by:
\begin{subequations}
    \begin{align}
        &\mu_1  =\sum_{i=1}^{N_{\mathrm{RF}}} s_{i,j} \sin \theta_i \sin \phi_i, \quad
        \eta_1  =\sum_{i=1}^{N_{\mathrm{RF}}} s_{i,j} \sin \theta_i \cos \phi_i ,\label{subs1}
        \\
        &\mu_2=\sum_{i=1}^{N_{\mathrm{RF}}} s_{i,j} \sin \theta_i \cos \big(\phi_i-\hat{\phi}_j\big),
        \  
        \eta_2=\sum_{i=1}^{N_{\mathrm{RF}}} s_{i,j} \cos \theta_i,\label{subs2}
    \end{align} 
\end{subequations}
where the calculation of~\eqref{phi} and~\eqref{subs1} are executed before~\eqref{theta} and~\eqref{subs2}, that is, $\hat{\phi}_j$ is known during the calculation of $\mu_2$.
Upon the derivation of the angular variables $\{\theta_j, \phi_j\}$, we then solve~\eqref{l-alpha} to obtain the concentration estimates, i.e., $\hat{\alpha}_j$ as follows
\begin{equation}\label{alpha}
    \operatorname{coth} \hat{\alpha}_j-\frac{1}{\hat{\alpha}_j}=\frac{\mu_3}{\eta_3},
\end{equation}
where $\mu_3, \eta_3$ are also substitution variables defined as
    $
        \mu_3=\sum_{i=1}^{N_{\mathrm{RF}}}s_{i,j}\left\{\sin \theta_i \sin \hat{\theta}_j \cos \big(\phi_i-\hat{\phi}_j\big)+\cos \theta_i \cos \hat{\theta}_j\right\},
        $ and $
        \eta_3=\sum_{i=1}^{N_{\mathrm{RF}}} s_{i,j}
        $.
(\ref{alpha}) is a transcendental equation that could be solved numerically via existing algorithms (e.g., Newton's iteration method). So far, the ML estimation of vMF distribution has been obtained.
\subsection{Wavenumber-Domain Sampling Method}\label{CMdesign}
The sampling procedure in Sec.~\ref{MLofVMF} implies that each observation signal behind each RF chain, i.e., $[{\bf y}]_i$ corresponds to a designated wavenumber-domain value $s_{i,j}$ at the angular direction of $\{\theta_i, \phi_i\}$.
This is achieved by designing the combining matrix ${\bf C}$ as follows:
\begin{equation}\label{C}
\begin{aligned}
        \mathbf{C}=\mathbf{B} \boldsymbol{\Psi}^H, \quad[\mathbf{B}]_{{i, |m|}}&=\left\{\begin{array}{l}
1, \text { if } {|m|}=\left\lfloor\frac{|\xi|}{{N}_{\rm RF}}\right\rfloor(i-1)+1 \\
0, \text { else }
\end{array}\right. 
,
\end{aligned}
\end{equation}
where $|m|$ is the sequence number for row-by-row or column-by-column scanning of the wavenumber domain indices.
Correspondingly, the combing matrix design in~\eqref{C} ensures each $i$-th entry in the received signal, i.e., $\vert[{\bf y}]_i\vert^2$ is the uniform sampled wavenumber domain value, i.e., $s_{i,j}$. This is because $[\mathbf{y}]_i\sim\mathcal{CN}(0,A^2(\theta_i,\phi_i))$.
We then define the $i$-th sampling wavenumber domain index as $m_i \triangleq (m_{x, i}, m_{y,i})$.
Considering the $i$-th wavenumber vector expression $k_{x,i}=\frac{m_{x,i}}{L_x}=k\sin\ \theta_i\cos\ \phi_i$, $k_{y,i}=\frac{m_{y,i}}{L_y}=k\sin\ \theta_i\sin\ \phi_i$ and $k_{z,i}=\sqrt{1- k^2_{x,i}- k^2_{y,i}}=k\cos\ \theta_i$, the relationship between the $(m_{x,i},m_{y,i})$-th wavenumber and its corresponding AoA and ZoA $\{\theta_i,\phi_i\}$ can be calculated by
\begin{equation}\label{thetai}\small
\begin{split}
\theta_i=
\left\{\begin{array}{l}
\arcsin\Big(c\frac{\sqrt{(\frac{m_{x,i}}{L_x})^2+(\frac{m_{y,i}}{L_y})^2}}{f_c}\Big),\ \ \ \ \ \ {\rm if}\ m_{x,i}m_{y,i}>0, \\ 
\\
\arcsin\Big(c\frac{\sqrt{(\frac{m_{x,i}}{L_x})^2+(\frac{m_{y,i}}{L_y})^2}}{f_c}\Big)+\pi,\ {\rm if}\ m_{x,i}m_{y,i}<0,
\end{array}\right.
\end{split}
\end{equation}
\begin{equation}\label{phii}\small
\begin{split}
\phi_i=
\left\{\begin{array}{l}
\arctan({m_{y,i}}/{m_{x,i}}),\ \ \ \ \ \ \ \ {\rm if}\ m_{x,i}>0,\ m_{y,i}>0, \\ 
\\
\arctan({m_{y,i}}/{m_{x,i}})+\pi,\ \ \ {\rm if}\ m_{x,i}<0, \\
\\
\arctan({m_{y,i}}/{m_{x,i}})+2\pi,\ \ {\rm if}\ m_{x,i}>0,\ m_{y,i}<0,
\end{array}\right.
\end{split}
\end{equation}
respectively.
Note that $(m_{x,i},m_{y,i})$ is restricted within (\ref{xi}), so the $\arcsin(\cdot)$ function in (\ref{thetai}) is always meaningful.

\subsection{Proposed WD-EM Algorithm}
Our goal is to acquire the statistical parameters of each $j$-th scatterer of non-zero entries, i.e., the weights $w_j$, the concentration parameters $\alpha_j$, the AoAs $\phi_j$ and the ZoAs $\theta_j$ as shown in~\eqref{vmf}. 
The problem formulated above is essentially a maximum likelihood (ML) estimation where the to-be-estimated parameters $\{w_j, \alpha_j, \theta_j, \phi_j\}_{j\in \{1,\dots, N_c\}}$ is determined to maximize the likelihood function $\mathcal{L}$ in \eqref{likelihood}.
However, traditional ML-based methods fail shortly since we need to determine which non-zero entry belongs to the cluster.
Thus, a novel wavenumber-domain expectation-maximization (WD-EM) algorithm is proposed, where the expectation step (E-step) categorizes the non-zero entries into their designated cluster; the maximization step (M-step) derives the ML estimation of the parameters given the categorization results in the E-step.
A more specific illustration is as follows:
\subsubsection{\textbf{E-step}}
As illustrated in Sec.~\ref{CMdesign}, each sample value $s_i$ is composed of $N_c$ components corresponding to $N_c$ clusters, i.e., $s_i = \sum_{j}^{N_c} s_{i,j}$. Thus, the categorization process is achieved by estimating $s_{i,j}$ given the parameter estimates in the M-step, i.e.,
\begin{equation}\label{sj}
    \hat{s}_{i,j} = \frac{p_{i,j}}{\sum_{j}^{\tilde{N}_c}p_{i,j}} s_i,
\end{equation}
 where $\tilde{N}_c$ is an input constant denoting the maximum number of the scatterers, which satisfies $\tilde{N}_c \geq N_c$. The weight estimate $\hat{w}_j$ of the redundant scatterer indices will be automatically calculated as zeros during iterations. $p_{i,j}$ is the likelihood probability function of $s_{i}$ at its corresponding sampling point $(\theta_i, \phi_i)$ to be generated by the $j$-th cluster, given by
\begin{equation}\label{pij}
    p_{i,j} = \hat{w}_j \cdot p(\theta_i, \phi_i; \hat{\alpha}_j, \hat{\theta}_j, \hat{\phi}_j),
\end{equation}
where $\{\hat{w}_j, \hat{\alpha}_j, \hat{\theta}_j, \hat{\phi}_j\}_{j\in \{1,\dots, \tilde{N}_c\}}$ are the estimates given by the {\it M-step}.
\subsubsection{\textbf{M-step}}
Given the categorization results, i.e., $\hat{s}_{i,j}, \forall i \in \{1, \dots, N_{\rm RF}\}, \forall j\in \{1,\dots, \tilde{N}_c\}$, the {\it M-step} derives the ML estimates of the parameters $\{\hat{w}_j, \hat{\alpha}_j, \hat{\theta}_j. \hat{\phi_j}\}_{j\in \{1,\dots, \tilde{N}_c\}}$.
Specifically, $\{\hat{\alpha}_j, \hat{\theta}_j. \hat{\phi_j}\}_{j\in \{1,\dots, \tilde{N}_c\}}$ can be calculated by \eqref{alpha}, \eqref{theta} and \eqref{phi}, respectively, while the weight estimates $\{\hat{w}_j\}_{j\in \{1,\dots, \tilde{N}_c\}}$ are given by
\begin{equation}\label{wj}
    \hat{w}_j = \frac{\sum_{i}^{N_{\rm RF}} \hat{s}_{i,j}}{\sum_{i}^{N_{\rm RF}}\sum_{j}^{\tilde{N}_c}\hat{s}_{i,j}}.
\end{equation}
The overall algorithm is summarized in~\textbf{Algorithm \ref{algorithmchart}}.

\begin{figure*}
    \centering
    \begin{minipage}{0.49\textwidth}
        \centering
        \includegraphics[width=0.8\linewidth]{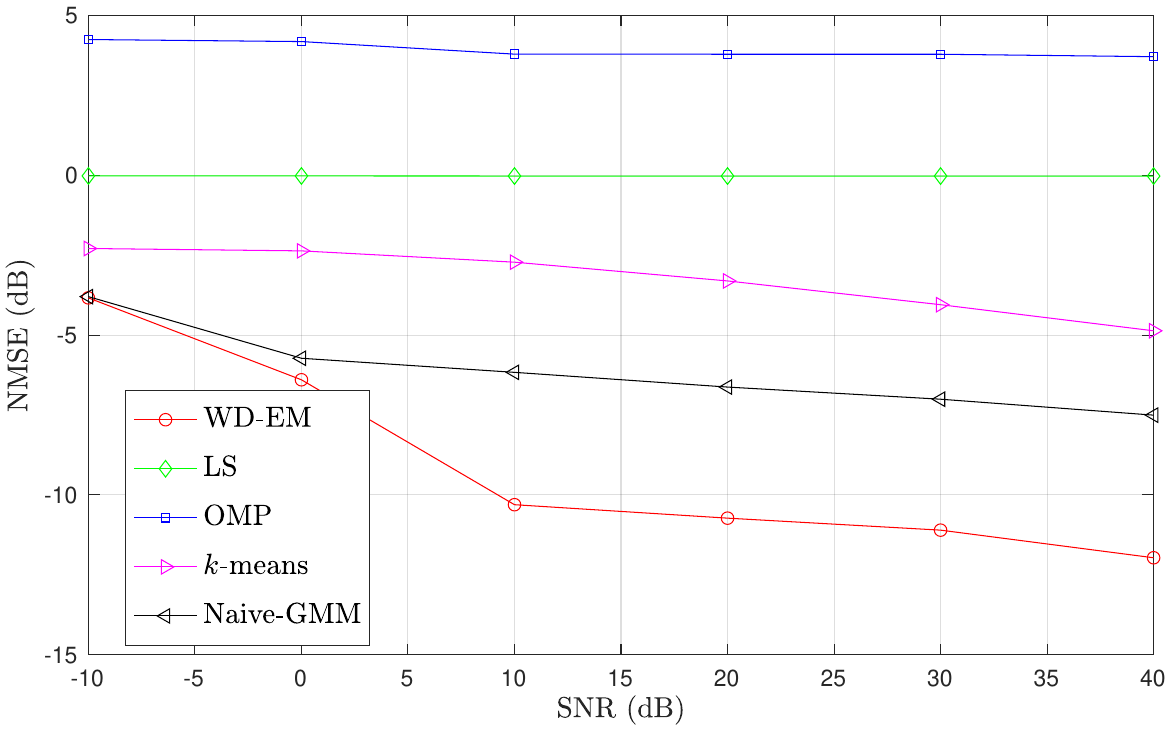}
        \caption{NMSE versus SNR, with $N_{\rm RF} = 200$.}\label{Sim1}
    \end{minipage}
    \hfill
    \begin{minipage}{0.49\textwidth}
        \centering
        \includegraphics[width=0.8\linewidth]{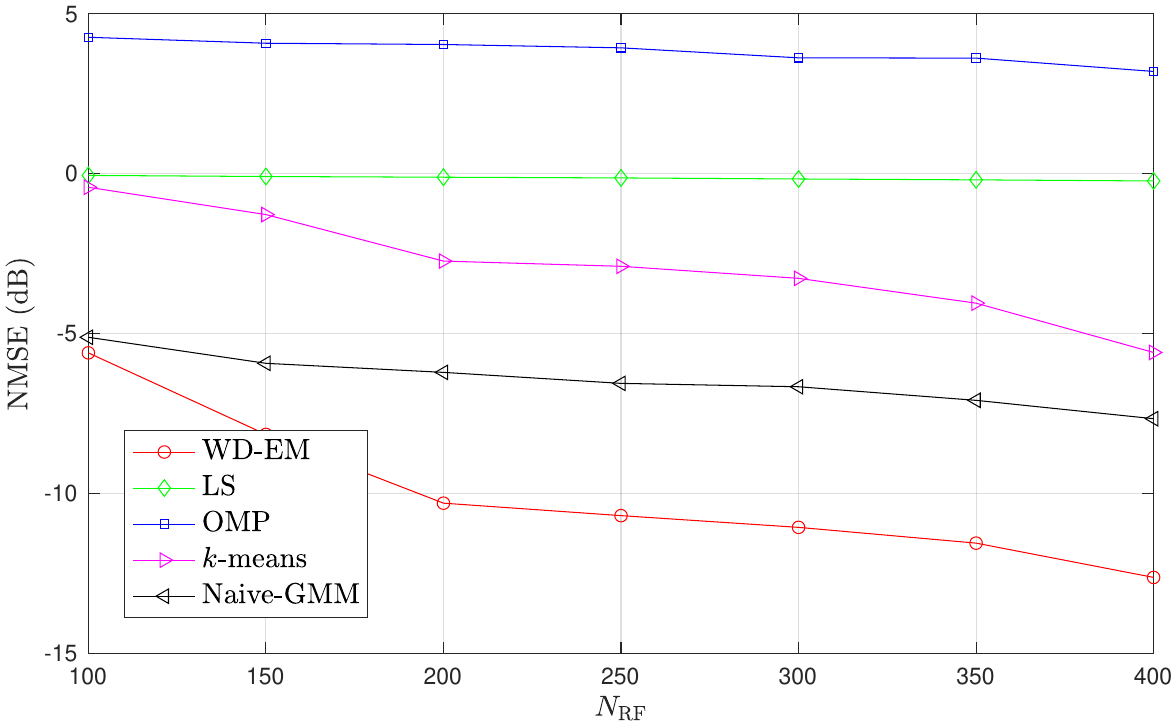}
        \caption{NMSE versus $N_{\rm RF}$, with ${\rm SNR} = {\rm 10\ dB}$.}\label{Sim2}
    \end{minipage}
    \vspace{-0.5cm}
\end{figure*}

\section{Simulation Results}\label{sec4}
Simulation results are presented in this section to evaluate the effectiveness of our proposed scheme. More specifically, the proposed cluster-sparsity-assisted sparse recovery algorithm, designated as WD-EM, is benchmarked against the commonly used least squares (LS) method, which assumes non-sparsity, as well as against other algorithms that exploit basic sparsity like the orthogonal matching pursuit (OMP), and leverage the clustered sparsity, such as the $k$-means and Naive GMM algorithms. The results demonstrate that our proposed algorithm significantly outperforms these benchmarks regarding robustness to signal-to-noise ratio (SNR) and the number of observations.

\subsection{Simulation Setup}
The carrier frequency is set to $30$~GHz. The size of the HMIMO is set to $1$~m$\times1$~m. The antenna spacing is set to $\delta=\lambda/2$, where $\lambda$ denotes the carrier wavelength. 
The maximum number of scatterers for WD-EM input is set to $\tilde{N}_c=4$, and the actual number of scatterers is set to $N_c=3$ to verify the robustness of the proposed algorithm. 
The parameters $\{w_j, \alpha_j, \theta_j, \phi_j\}_{j\in \{1,\dots, N_c\}}$ are randomly generated following the uniform distributions $\mathcal{U}(0, 1)$, $\mathcal{U}(50, 100) $, $\mathcal{U}(0, \pi/2)$, and $\mathcal{U}(0, 2\pi)$, respectively.
The signal-to-noise rate (SNR) is defined as $\frac{\vert\vert\mathbf{CH}x\vert\vert^2}{N_{\rm RF}\sigma_0^2}$.
We choose the normalized mean square error (NMSE) as the simulation metric, given by
\begin{equation}\label{nmse}
{\rm NMSE}={\mathbbm{E}\{\vert\vert \mathbf{R_H}-\mathbf{\hat{R}_H}\vert\vert^2_{\it{F}}\}}/{\mathbbm{E}\{\vert\vert\mathbf{R_H}\vert\vert^2_{\it{F}}\}},
\end{equation} 
where $\mathbf{R_H}$ denotes the covariance of the channel matrix
\begin{equation}\label{covariance}
    \mathbf{R_H}=\mathbbm{E}\{\mathbf{HH^H}\}=\mathbf{\Psi}\mathbbm{E}\{\mathbf{ss^H}\}\mathbf{\Psi^H}=\mathbf{\Psi}\mathbf{D}\mathbf{\Psi^H},
\end{equation}
where $\mathbf{D}\in\mathbbm{R}^{\vert\xi\vert\times\vert\xi\vert}$ is a diagonal matrix, and its diagonal entries collect all $\vert\xi\vert$ variances, e.g. $[\mathbf{D}]_{ii}=\sigma^2(m_{x,i},m_{y,i})$. 
We adopt the following benchmarks:
\begin{itemize}
        \item \textbf{WD-EM}: The proposed WD-EM algorithm demonstrated in $\mathbf{Algorithm\ 1}$. 
    \item \textbf{LS}: Estimating the whole channel without sparsity using least-square (LS) estimation.
    \item \textbf{OMP}~\cite{omp}: Taking the most basic sparsity and solving non-zero entries using entry-by-entry matching persuit.
    \item \textbf{${\boldsymbol k}$-means}~\cite{clusteringalgorithms}: Group the non-zero elements using the classical $k$-means algorithm and calculate the non-zero values using pseudo-inverse.
    \item \textbf{Naive-GMM} \cite{clusteringalgorithms}: Unlike our proposed GMM-based WD-EM algorithm, this benchmark only groups the non-zero entries using GMM-based modeling, without the ability to estimate the weight factors $w_j$. 
\end{itemize}


\subsection{Performance Evaluation}
Fig.~\ref{Sim1} illustrates the NMSE versus SNR with $N_{\rm RF}=200$. As anticipated, the classic entry-wise LS method exhibits significant estimation errors due to the extreme dimensionality of the HMIMO channel.
The NMSE of the OMP-based CS method remains above 0 dB. This is because the number of non-zero entries in $\mathbf{H}$ (denoted as $K$) significantly exceeds the traditional CS-based recovery threshold, i.e., $2K < N_{\rm RF}$~\cite{CS}, resulting in the poorest performance.
The $k$-means method fails to compete with the Naive-GMM and our proposed WD-EM methods since it only exploits the location information of the non-zero entries while neglecting the value information.
Both the Naive-GMM and the WD-EM methods effectively leverage the values and locations of the non-zero entries for grouping. However, the proposed WD-EM method outperforms the Naive-GMM by up to 5 dB, as the Naive-GMM fails to accurately determine the weight factors, as shown in~\eqref{wj}.

Fig.~\ref{Sim2} illustrates the NMSE versus $N_{\rm RF}$, representing the number of observations. 
As depicted, all benchmarks experience notable performance degradation, except for the proposed WD-EM method, as the number of observations decreases from $N_{\rm RF} = 400$ to $N_{\rm RF} = 100$, 
This is because our proposed WD-EM algorithm efficiently compresses the dimensions of the parameters to be estimated from a number proportional to the array aperture to the order of the scatterer number. Consequently, the WD-EM method requires significantly fewer observations compared to others, resulting in strong robustness against reductions in $N_{\rm RF}$.

\section{Conclusion}\label{sec5}
This paper investigates the sparse channel estimation for HMIMO systems leveraging the clustered sparsity.
We first unmask the clustered structured inherent in the locations and values of the non-zero entries within the wavenumber domain channel. 
Initially, we unveiled the inherent clustered structure residing in both the locations and values of non-zero entries within the wavenumber domain channel. 
Upon this insight, we adopt a GMM-based probabilistic model to effectively capture such clustered sparsity.
Through this approach, we managed to compress the dimensions of the parameters to be estimated significantly down to the order of the scatterer number.
Finally, a novel WD-EM algorithm is proposed to derive the precise posterior channel parameters.
With the capability to capture clustered sparsity and achieve remarkably low-dimensional complexity, our proposed WD-EM algorithm outperforms existing benchmarks in terms of robustness against SNR variations and the reduction of RF chains.


\IEEEpeerreviewmaketitle

\vspace{-0.2cm}
\bibliographystyle{IEEEtran}
\bibliography{WCL-GYQ} 

\end{document}